\documentclass[prb,aps,preprint,showpacs,showkeys,superscriptaddress,reprint,amsmath,amssymb]{revtex4-1}
\usepackage{graphicx}% Include figure files  
\usepackage{color}
\usepackage{amssymb,amsmath}
\usepackage{comment}

\usepackage{hyperref}
\hypersetup{
bookmarks=false,
pdftitle={coupled cMPS},
anchorcolor=blue,
colorlinks=true,
citecolor=blue,
urlcolor=blue,
linkcolor=blue}
\usepackage{xcolor}
\usepackage{bbold}

\begin{document}

\title{Continuous matrix product states  for  coupled fields:
\\
  Application to Luttinger Liquids and quantum simulators}

\author{Fernando Quijandr\'{\i}a} 
\affiliation{Instituto de Ciencia de Materiales de Arag\'on y Departamento de F\'{\i}sica de la Materia Condensada, CSIC-Universidad de Zaragoza, Zaragoza, E-50012, Spain.}

\author{Juan Jos\'e Garc\'{\i}a-Ripoll}
\affiliation{Instituto de F\'{\i}sica Fundamental, IFF-CSIC, Serrano 113-bis, Madrid E-28006, Spain}

\author{David Zueco} 
\affiliation{Instituto de Ciencia de Materiales de Arag\'on y Departamento de F\'{\i}sica de la Materia Condensada, CSIC-Universidad de Zaragoza, Zaragoza, E-50012, Spain.}
\affiliation{Fundaci\'on ARAID, Paseo Mar\'{\i}a Agust\'{\i}n 36, Zaragoza 50004, Spain}

\date{\today}

\begin{abstract}
A way of constructing continuous matrix product states (cMPS) for
coupled fields is presented here.  
The cMPS is a variational \emph{ansatz}
for the ground state of quantum field theories in one dimension.  
Our proposed scheme is based in the physical interpretation in which the cMPS class can be produced by means of a dissipative dynamic of a system interacting with a bath.
 We study the case of coupled bosonic fields.
 We test the method  with previous DMRG results in coupled Lieb Liniger models.
Besides, we discuss a novel application for characterizing the
Luttinger liquid theory emerging in the low energy regime of these theories.
Finally, we propose a circuit QED
architecture as a quantum simulator for coupled  fields.
 \end{abstract}

\pacs{Valid PACS appear here}% PACS, the Physics and Astronomy
                             % Classification Scheme.
%\keywords{Suggested keywords}%Use showkeys class option if keyword
                              %display desired
\maketitle

%\tableofcontents

%%%%%%%%%%%%%%%%%%%%%%%%%%%%%%%%%%%
%%%%%%%%%% intro %%%%%%%%%%%%%%%%%%%%%
%%%%%%%%%%%%%%%%%%%%%%%%%%%%%%%%%%%

\section{Introduction}

%%%%% q-info -> many body 
Quantum Information and Quantum Technologies are providing both a new
language and a new experimental landscape for the study of large
quantum many-body systems. The study of
entanglement in extended lattice models has made it possible to tackle
the successful numerical renormalization group (NRG) \cite{Wilson75}
and the density matrix renormalization group  (DMRG)  \cite{White92,Schollwoeck05} and provide them with a solid
theoretical background based on the distribution of bipartite
entanglement in 1D systems. This understanding made it possible to
introduce new methods based on the matrix product states   (MPS) formalism that allow studying
both static\ \cite{Porras04} and time-dependent phenomena\
\cite{Vidal04,Verstraete04,Feiguin04,Daley04,Garcia-Ripoll06},
together with generalizations for critical\ \cite{Vidal07} and
two-dimensional systems\ \cite{Verstraete08}.
As examples of the success of these methods we can remark the
extremely good accuracy of DMRG studies in studying quantum phase
transitions of lattice models\ \cite{Schollwoeck05}, as well as the
success in the quantitative modelisation of novel experiments with
cold atoms\ \cite{Cheneau12,Trotzky12,Fukuhara13}, ions\ \cite{Hauke13} and
photonic systems\ \cite{Peropadre14,Burillo}.

%%%%%% qft -> cmps

The above examples rely on lattice models.  Sometimes, however,
physics is best described via continuum 1D field theories. This
includes 1D Bose-Einstein condensates under strong confinement and
interaction, long Josephson junctions or nonlinear
materials\cite{giamarchi, Cazalilla2011,
  Steffens2014,Steffens2014b,Drummond2014}. In the seminal work of Verstraete and Cirac
\cite{Cirac-Verstraete}  the MPS formalism was extended to treat 
 continuum 1D quantum mechanical systems. The continuous
matrix product state (cMPS) was formulated as a variational {\it
  ansatz} for obtaining ground states of continuum one dimensional and
non relativistic fields \cite{Haegeman}.  More recently, the cMPS
formalism has been used for tackling excited (1 particle) states
\cite{Draxler2013} and $1 + 1$ relativistic theories
\cite{Haegeman2010, Stojevic2014}.

%%%%%% our work -> coupled fields 

In this work we introduce a natural extension of cMPS states to
study coupled fields. We show that, thanks to the cMPS formalism, the
interaction between fields does not have to be treated perturbatively,
developing the appropriate algorithms to compute ground state
properties. This is the main result of the paper and, by itself, it
has potential applications to describe systems present experiments
with interacting 1D Bose gases\cite{Cazalilla3}, as well as the fermion-fermion or
fermion-boson interactions, occurring in the so-called ladders\cite{giamarchi, Cazalilla2011}.

%%%%%%%%luttinger liquids.

A well known peculiarity for one dimensional models
is their low-Energy description as {\it Luttinger liquids}
\cite{Haldane1, Haldane2}.  These liquids, no matter of the original
model, are effective theories of bosonic character and are described
by {\it Sine-Gordon}-like models.  The parameters in the effective
theory must be extracted from the original (microscopic) model.  In
the case of field theories, the parameters are given in terms of the
ground state \cite{Cazalilla1, Orignac2010}.  The second important
result in this work is that cMPS can be used to derive those Luttinger
parameters, both for the single and the coupled field case.

%%%%%%%%%% holographic -> q-simulators.

Finally, we also relate the coupled cMPS ansatz to the
simulation of coupled quantum fields. 
We 
provide a recipe for building cMPS in the lab: engineering discrete
quantum systems coupled to transmission lines, as in
circuit QED setups.  Those lab-layouts are nothing but
prototypes for quantum simulators of field theories within cavity QED
\cite{Barrett2013}.

%%%%%%%%%%%% final paragraph:  organization of the paper

The rest of the paper is organized as follows. The following section
 \ref{sec:ovcmps} is an (almost) self contained summary of the
cMPS theory.  Next, Sect. \ref{sec:LL} is our first application.  We
use the cMPS, still single field, for obtaining the parameters in the
Luttinger liquid theory, explained in \ref{sec:boso} and applied to
the Lieb Linniger model, Subsect. \ref{sec:calc-LL}.  
section \ref{sec:extension} explains our extension for coupled fields
and \ref{sec:coupledLL} reports in our numerical results for coupled
bosonic species.  We finish, in section \ref{sec:disc}, commenting on the application of cMPS for
constructing quantum simulators and summarize our results.

%%%%%%%%%%%%%%%%%%%%%%%%%%%%%%%%%%%%%%%%
%%%%%%%%%%%%%%%% overview  %%%%%%%%%%%%%%%%%
%%%%%%%%%%%%%%%%%%%%%%%%%%%%%%%%%%%%%%%%

\section{Overview of cMPS}\label{sec:ovcmps}

%%%%%%%%%%% figure sketch %%%%%%%%%%%%%
\begin{figure}[t]
\includegraphics[width=1.\columnwidth]{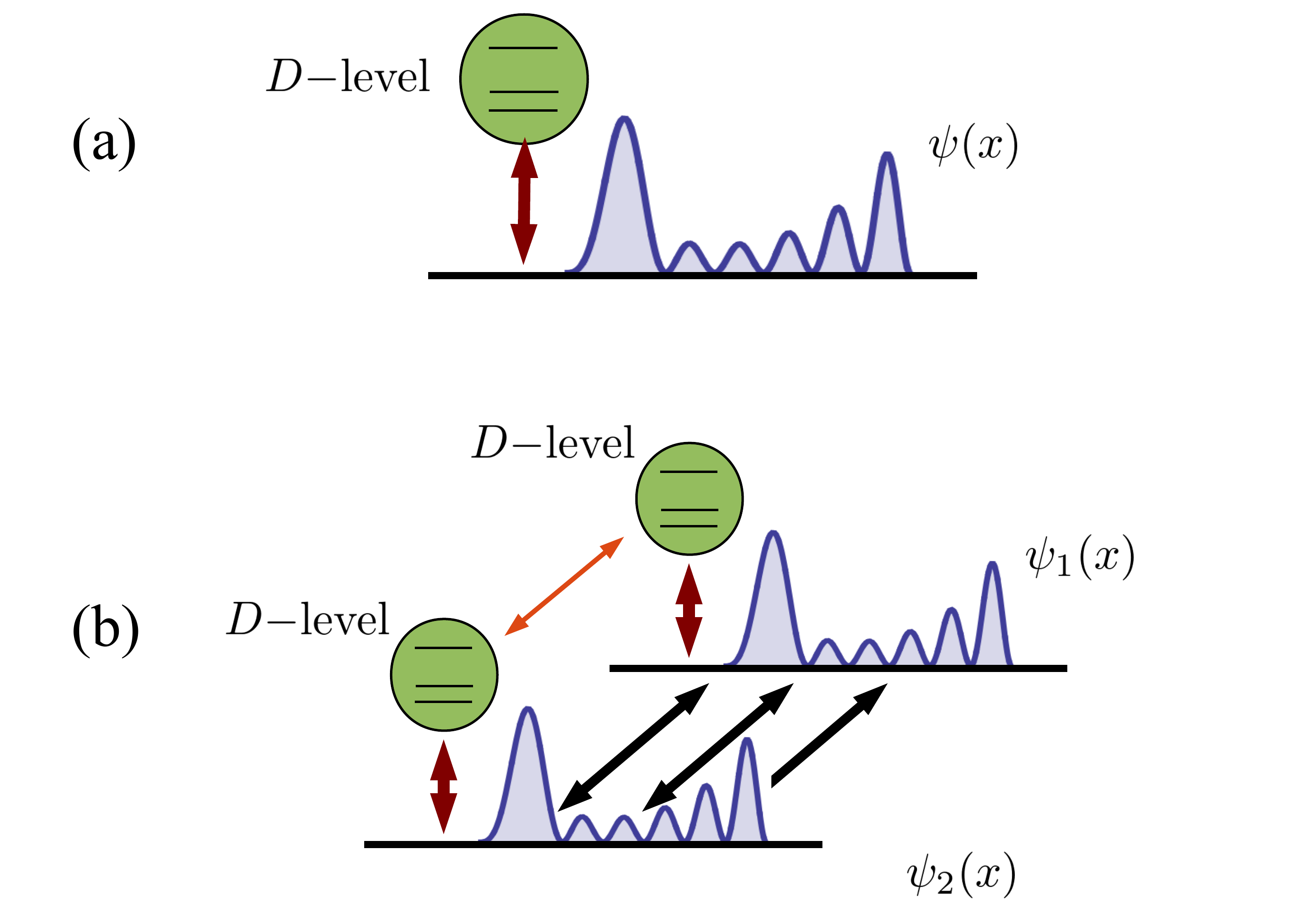}
 \caption{(color online) (a) The cMPS can be built by letting an ancilla (green circle in
the figure) be coupled to a continuous field ($\psi(x)$). (b) In a similar
way, we propose that coupled fields can also be constructed by coupling independent
ancillas.}
 \label{fig:sketch}
 \end{figure}

%%%%%%%%%%%%%%%%%%%%%%%%%%%%%%%%%%%
We review here the basics of continuous matrix product states (cMPS) for a single  
field.
The cMPS are trial states for a
variational estimation of ground states in one-dimensional quantum field theories.
We start by considering a quantum system described in second
quantization by means of the field operators
$\hat{\psi}(z)$. According to the spin-statistics theorem, these
operators must satisfy (anti)commutation relations 
$\hat{\psi}(z) \hat{\psi} ^\dagger(z^\prime) \pm  \hat{\psi} ^\dagger(z^\prime) \hat{\psi}(z) =\delta(z-z^\prime)$ according to whether they are fermions or bosons. Our system will be defined in a length $L$ with periodic boundary conditions.

The explicit form of the state can be written, as introduced in the
seminal work of Verstraete and Cirac \cite{Cirac-Verstraete} ($\hbar =
1$ is used through the text) 
\begin{equation}
\label{cMPS-def}
\vert \chi \rangle = {\rm tr_{aux}} \lbrace\mathcal{P} {\rm e}^{
  \int_0^L {\rm d}z \, ( Q(z)\otimes \mathbb{1}_{\mathcal F} + R(z)\otimes
  \hat{\psi}^{\dagger}(z) ) } \rbrace \vert \Omega \rangle
\; 
\end{equation}
where $\mathcal{P}$ denotes path-ordering (we follow the prescription in which, for  
the argument of the exponential, the value of $z$ increases as we move to the right).
$Q(z)$ and $R(z)$ are complex $D \times D$ matrices acting on an
auxiliary Hilbert space 
$\mathcal H_{\rm aux}$.
The partial trace  ${\rm
  tr_{aux}}$ is taken over $\mathcal H_{\rm aux} $.  
The suffix in $\mathbb{1}_{\mathcal F}$ 
emphasizes that it is the identity for the field. 
Finally, the state $|\Omega\rangle$ is the vacuum of a free theory,
\begin{equation}
\label{vac}
\hat{\psi}(z) | \Omega \rangle = 0
\end{equation} 

From now on, we will restrict ourselves to translational invariant setups in which the matrices $R$ and $Q$ become independent of $z$. 
The cMPS are complete \cite{Brockt2012}, {\it i.e.}, any one dimensional quantum field can
be casted in the form \eqref{cMPS-def}. 
This class of states can be
obtained as the continuum limit of MPS, with bond dimension $D$.   
The bond dimension can be understood as a measure of the block
entanglement.  In one dimension the block entanglement saturates,
thus, $D$ is expected to be sufficiently small. 
If so, we are able to reach any quantum state with a relatively small number of variational parameters ($2D^2$). This combined with the variational method results in a very powerful technique for finding ground-states of one-dimensional theories.

In a relativistic scenario, the block-entanglement  has a UV logarithmic divergence.
This can be understood since the  ground state of a relativistic theory, also in $1+1$, will contain
zero-point fluctuations from all energy scales, which are the ones
contributing more to the entropy \cite{Srednicki1993}.  
A related argument due to Feynman is quoted  in
Ref. \onlinecite{Haegeman2010}.  
The ground state will be dominated by the high energy contributions.  
As a consequence, in the variational procedure, the accuracy for
describing the low-E sector is lost.
Therefore, a cutoff must be introduced.
Though challenging, the description of relativistic field theories has
been succesfully described via cMPS introducing a regularization scheme
\cite{Haegeman2010, Stojevic2014}.
 Here we will face the most favourable case of one dimensional non-relativistic theories.

Inherited from their discrete countenparts, the cMPS is not unique but
the {\it gauge} transformation $Q \to g Q g^{-1}$ and $R \to g R g^{-1}$
leaves the state $| \chi \rangle$ invariant \cite{Haegeman2013}.
It turns out that the gauge 
\begin{equation}
\label{gauge}
Q + Q^\dagger + R^\dagger R=0
\end{equation}
is quite convenient.  In this gauge,
the cMPS state 
(\ref{cMPS-def}) can be rewritten as, 
\begin{equation}
\label{cMPS-U}
| \chi \rangle = {\rm tr_{aux}} 
\{
U(L,0)
\}
|\Omega \rangle
\end{equation}
with
\begin{equation}
\label{U}
U(L,0) = {\mathcal P}{\rm e}^{ -i \int_0^L {\rm d}z (K \otimes  \mathbb{1}_{\mathcal F}  +i R \otimes \hat{\psi}^\dagger(z)
-i  R^\dagger \otimes \hat{\psi} (z))}
\end{equation}
and, $K=K^\dagger$ Hermitean:
\begin{equation}
\label{Q-single}
Q = - i K - \frac{1}{2} R^\dagger R 
\; ,
\;
\end{equation}
which implies the gauge condition
(\ref{gauge}).
The unitary operator $U$, in Eq. \eqref{U},  is  formally equivalent to a
evolution in $z$-time for the field $\hat{\psi} (z)$ and a $D$-level (auxiliary) system with Hamiltonian $K$.
Field and ancilla are coupled via $ i R \otimes \hat{\psi}^\dagger (z) - i R^{\dagger} \otimes \hat{\psi}(z)$.
The ground state is described in terms of the matrices $K$ and $R$,
{\it i.e.}, in terms of an auxiliary zero-dimensional system. This suggests an
holographic interpretation for the cMPS  \cite{Eisert}. See Fig.
\ref{fig:sketch}(a) for a pictorial interpretation. 

It remains to
provide operational rules for computing within the cMPS formalism.
To be precise, we must be able to write any field observable 
$\langle \chi | O (\hat{\psi}, \hat{\psi}^\dagger) | \chi \rangle $ in terms of the
matrices $R$ and $Q$.
As detailed in Ref. \onlinecite{Rispler2012},
the following  relations are found:
\begin{align}
\label{norm-0}
\langle \chi | \chi \rangle &= {\rm tr}  \{ {\rm e}^{T L} \}
\\ 
\langle \hat{\psi} ^\dagger (z ) \hat{\psi} (z) \rangle
&= {\rm tr}  \{ {\rm e}^{T L} ( R \otimes R^* )  \}
\\ 
\langle \partial_z \hat{\psi}^{\dagger}(z) \partial_z \hat{\psi}(z) \rangle &= {\rm tr} \{ {\rm e}^{TL} ([Q,R]\otimes [Q^*,R^*]) \}
\\
\langle \hat{\psi}^{\dagger}(z) \hat{\psi}^{\dagger}(z) \hat{\psi}(z) \hat{\psi}(z) \rangle &= {\rm tr} \{ {\rm e}^{TL} (RR \otimes R^*R^*) \}
\\
\nonumber
\end{align}
here,
\begin{equation}\label{transfer1}
T = Q \otimes \mathbb {1}
+
\mathbb {1} \otimes Q^* + R \otimes R^*
\end{equation}
The Kronecker products in the ancilla space occurs because
some identities,
{\it e.g.},
${\rm tr} \{A \} {\rm tr} \{B \} ={\rm tr} \{A  \otimes B\}$, have been
used.

To avoid those products in the
auxiliary space, the isomorphism $|a \rangle | b \rangle \to |a
\rangle \langle b^* |$ is introduced. This allows us to map vectors in $\mathcal H_{\rm aux} \otimes \mathcal H_{\rm aux} $ into operators acting on $\mathcal H_{\rm aux}$. This can be understood from the fundamental property
\begin{align}
%\label{iso-trace}
\nonumber
&{\rm tr}
%_{\mathcal H \otimes \mathcal H} 
 \Big \{ 
\sum_i \sum_{a b c d } A_i \otimes B_i^* |a \rangle |b \rangle \langle
c | \langle d |
\Big \}
=
\\ 
%\nonumber
\label{iso-trace}
&{\rm tr}
%_{\mathcal H}
\Big \{ 
\sum_i \sum_{a b c d }
|d^* \rangle \langle c | A_i |a \rangle \langle b^* | B_i^\dagger 
\Big \}
\, .
\end{align}
The former also implies that operators acting on $\mathcal H_{\rm aux}$ are mapped into superoperators. Therefore, the action of $T$ on a ket $\vert \rho \rangle$ will be mapped into $\mathcal{T}[\rho]$, where $\mathcal{T}$ is a superoperator acting on the state (matrix) $\rho$. Under the isomorphism, it is straightforward to show that  
\begin{equation}
\label{Lind}
\mathcal{T}[\varrho]
:= 
-i [K, \varrho(z)]
+
R \varrho R^\dagger - \frac{1}{2} [ R^\dagger R, \varrho(z)]_+
\end{equation}
This is nothing but the dissipator governing  a Linblad-like evolution ${\rm d}_z \varrho = \mathcal T \varrho$ for the irreversible
dynamics of a system  coupled to a reservoir.
In this case,  the role of the system is being played by the ancilla and that of the bath by the field (see Fig. \ref{fig:sketch}(a) and the discussion above on the holographic
interpretation).
 The Linbladian is a positive-semidefinite operator,
$\mathcal T \leq 0$, having at least one zero eigenvalue
\cite{Rivas2011}. With this at
hand, Eq. \eqref{norm-0} can be rewritten as
\begin{equation}
\nonumber
\langle \chi | \chi \rangle
=
\langle l  | {\rm e}^{T L} r \rangle
=
{\rm Tr}  \Big ( {\rm e}^{T L} | r \rangle \langle l | \Big ) = 
{\rm Tr}  \Big ( l \cdot {\rm e}^{\mathcal{T} L} r
\Big )  
\; .
\end{equation} 
Here, $\langle l \vert$ and $\vert r \rangle$ are the left and
right eigenvectors of $T$ (respectively) associated with its zero
eigenvalue. We have assumed implicitly the limit $L \rightarrow
\infty$ where this eigenvalue yields the principal contribution to the
exponential. In the third equality, the above introduced isomorphism
has been used. Note that the zero eigenvectors of $T$, under the
isomorphism, are mapped into the stationary solutions of the Linblad
equation (left and right equations).
Accordingly, the action of $T$ into the bra $\langle l \vert$ can also be mapped into the action of a superoperator on a matrix: $ \langle l \vert T \Leftrightarrow Q^{\dagger} l + \varrho_l Q + 
R^{\dagger} l R $. It is easy to see that, under the gauge (\ref{gauge}), $l^* = \mathbb 1$ is a solution of the stationary Linblad-like dynamics ($ d_z l^* =0$).  \color{black}
Combining all of this, we end up with
\begin{equation}
\langle \chi | \chi \rangle =  {\rm tr}  ( r^*  ) = 1  
\end{equation}
where $ d_z r^* =0$.

In a similar way, we can re-express the expectation value of any operator in terms of the steady-state solution $\varrho^*$ of the right Linblad equation
\begin{align}
\label{av-1}
\langle \hat{\psi}^{\dagger}(z) \hat{\psi}(z) \rangle =& {\rm
  tr}\left(R^{\dagger}R\varrho^*\right) \\
\label{av-2}
\langle \partial_z \hat{\psi}^{\dagger}(z) \partial_z \hat{\psi}(z) \rangle =&
{\rm tr}\left(  [Q,R]^{\dagger} [Q,R] \varrho^* \right) \\
\label{av-3}
\langle \hat{\psi}^{\dagger}(z) \hat{\psi}^{\dagger}(z) \hat{\psi}(z) \hat{\psi}(z) \rangle =& {\rm tr}\left(  (R^{\dagger})^2 R^2 \varrho^*\right)
\end{align}
With this we conclude our overview of the cMPS formalism.  In the limit $L \rightarrow \infty$, we will be concerned with the ground state energy density $e_0 = \langle \chi \vert \hat{{\mathcal H}}(\psi , \psi^\dagger) \vert \chi \rangle$ (where $\hat{{\mathcal H}}$ is the Hamiltonian density operator). The latter can be computed by minimizing with the matrices $R$ and $Q$ as input and using the latter relations
(and similar ones).  Once the minimization procedure has
finished, observables can be computed with the same relations using the optimized matrices.

%%%%%%%%%%%%%%%%%%%%%%%%%%%%%%%%%%%%%%%%
%%%%%%%%%%%%%%%% app to LL  %%%%%%%%%%%%%%%%%
%%%%%%%%%%%%%%%%%%%%%%%%%%%%%%%%%%%%%%%%
\section{Application to Luttinger liquids}
\label{sec:LL}

%%%%%%%%%%%%%%%%%%%%%%%%%%%%%%%%%%%%%%%%
%%%%%%%%%%%%%%%% bosonization overview  %%%%%%%%%%%%%%%%%
%%%%%%%%%%%%%%%%%%%%%%%%%%%%%%%%%%%%%%%%
\subsection{Bosonization}\label{sec:boso}

At low temperatures, a large class of one dimensional theories exhibit excitations of bosonic nature and their correlation functions are characterized by power laws. An interesting feature of 1D is that this class makes almost no distinction between bosons and fermions. Haldane\cite{Haldane1,Haldane2}termed this class of theories {\it Luttinger liquids}. The bosonic nature of the low-energy excitations in 1D is due to the enhanced role quantum fluctuations acquire in low dimensional systems. 

For a given microscopic model, the so-called {\it bosonization} prescription, consists in expressing the original degrees of freedom in terms of new fields which capture the collective behaviour characterizing the low-energy regime. For the case of a bosonic field, we will introduce the density-phase representation\cite{giamarchi}
\begin{equation}
\hat{\psi}^{\dagger}(x) = \sqrt{\hat{\rho}(x)}{\rm e}^{-i \hat{\theta}(x)}
\end{equation} 
where $\hat{\rho}(x) := \psi^\dagger(x) \psi (x)$ is the particle density field and $\hat{\theta}(x)$ the phase field. Close enough to the ground state we can safely approximate the density operator by 
\begin{equation}\label{density-expansion}
\hat{\rho}(x) \sim \rho_0 - \frac{1}{\pi} \partial_x \hat{\phi}(x) 
\end{equation}
where $\rho_0$ is the ground state density and the operator $\hat{\phi}(x)$ characterizes the fluctuations over the ground state. The commutation relations for bosonic fields will translate into a canonical commutation relation for the $\hat{\theta}$ and $\hat{\phi}$ fields
\begin{equation}
[\frac{1}{\pi}\partial_x \hat{\phi}(x),\hat{\theta}(x')] = -i\delta(x-x')
\end{equation}

%%%%%%%%%%%%%%%%%%%%%%%%%%%%%%%%%%%%%%%%
%%%%%%%%%%%%%%%% LL parameters single field  %%%%%%%%%%%%%%%%%
%%%%%%%%%%%%%%%%%%%%%%%%%%%%%%%%%%%%%%%%

\subsection{Calculation for the Lieb-Liniger model}\label{sec:calc-LL}

We are going to apply the previous ideas to the Lieb-Liniger model \cite{Lieb-Liniger1}. The former describes a $1$D non-relativistic bosonic gas interacting via a repulsive zero-range potential

\begin{equation}\label{LL}
\hat{H} = \int_0^L {\rm d}x \, \frac{1}{2M} \partial_x \hat{\psi}^{\dagger}(x) \partial_x \hat{\psi}(x) + c \, ( \hat{\psi}^{\dagger}(x) )^2 ( \hat{\psi}(x) )^2
\end{equation}

The Lieb-Liniger model is exactly solvable by means of a Bethe {\it ansatz}. In fact, the solution shows that at low-energies, this model displays a Luttinger liquid behaviour \cite{Lieb-Liniger2}. An excellent agreement between the exact ground state energy density and the cMPS solution has already been provided \cite{Cirac-Verstraete}. Finally, note that this model conserves the particle number density. This quantity will represent a minimization constraint when finding the ground state numerically. 

Following the bosonization scheme, the effective Hamiltonian describing the low-energy behaviour of the Lieb-Liniger model is 

\begin{equation}
\hat{H}_{\rm eff} = \frac{v}{2\pi} \int_0^L {\rm d}x \, K (\partial_x \hat{\phi}(x))^2  + \frac{1}{K} (\partial_x \hat{\theta}(x))^2
\end{equation}

Hence, the low-energy regime can be completely characterized by means of two parameters (Luttinger parameters): the velocity $v$ and the dimensionless parameter $K$. These, in turn, can be related to the ground state energy density $e_0 (\rho)$ of the microscopic Hamiltonian (\ref{LL}). The corresponding relations are \cite{Cazalilla1} 

\begin{equation}\label{Lutt-v-single}
v^2 = \frac{\rho_0}{M} \left. \frac{\partial^2 e_0}{\partial \rho^2} \right\vert_{\rho = \rho_0} 
\end{equation}

\begin{equation}\label{Lutt-K-single}
K^2 = \frac{\pi^2 \rho_0}{M} \left( \left. \frac{\partial^2
      e_0}{\partial \rho^2} \right\vert_{\rho = \rho_0} \right)^{-1}
\; .
\end{equation}

\begin{figure}[t]
\includegraphics[width=1.0\columnwidth]{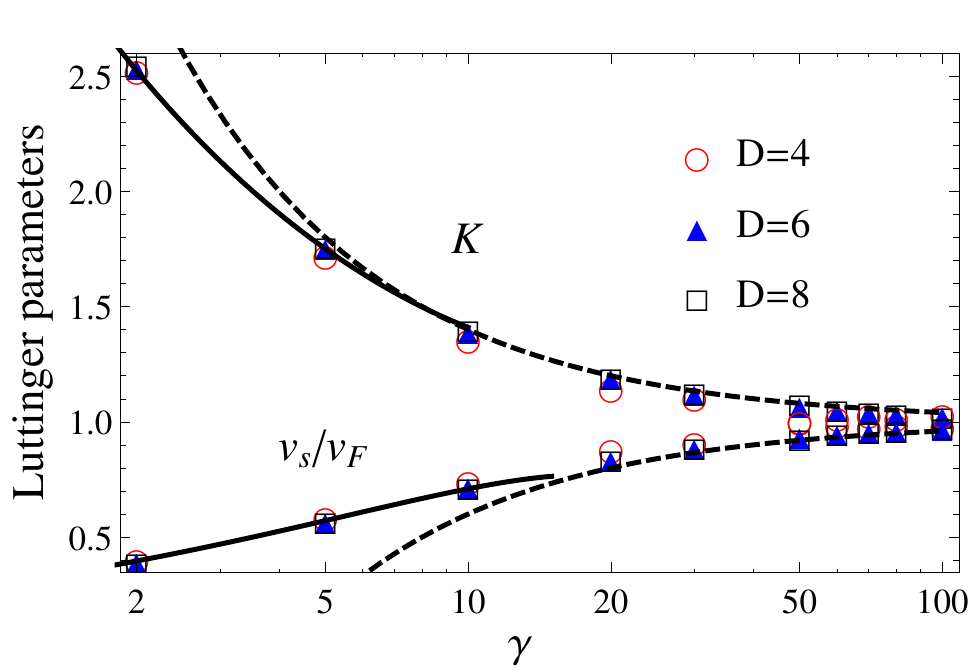}
 \caption{(color online) Luttinger parameters as a function of the dimensionless zero-range interaction constant $\gamma = Mc/\rho_0$ (we choose $\rho_0 =1$ and $M=1/2$). The full and dashed lines correspond to the weak and strong repulsion limits (respectively). These are compared with our cMPS results for bond dimensions $D= 4$ (open circles), $6$ (filled triangles) and $8$ (open squares). }
 \label{fig:Luttinger1}
 \end{figure}

It is possible to obtain asymptotic (analytic) expressions for the former parameters in terms of the dimensionless coupling constant $\gamma = M c /\rho_0$. In Fig. \ref{fig:Luttinger1} we compare those asymptotic limits (small and large repulsion, see Ref. \onlinecite{Cazalilla1}) for $v$ and $K$ with (\ref{Lutt-v-single}) and (\ref{Lutt-K-single}) as obtained from the ground state energy density computed with cMPS. We have performed simulations for $D=2$, $4$, $6$ and $8$. For every bond dimension, the ground state energy density was calculated for up to twelve different densities. By interpolation, we constructed the continuous function $e_0(\rho)$ and the derivatives were calculated from it.
Results show that for a moderately small bond dimension ($D=6$), it is possible to match the predicted asymptotic behaviour up to high values of $\gamma$. Results for $D=2$ are not shown for the sake of clarity (such a small bond dimension does not capture correctly the ground state of the Lieb-Liniger model).

%%%%%%%%%%%%%%%%%%%%%%%%%%%%%%%%%%%%%%%%
%%%%%%%%%%%%%%%% extension  %%%%%%%%%%%%%%%%%
%%%%%%%%%%%%%%%%%%%%%%%%%%%%%%%%%%%%%%%%

\section{Extension/Generalization for coupled fields}
\label{sec:extension}

The cMPS formalism can be naturally extended to treat a multi-species system. Let us consider a system of length $L$ in which coexist $q$ bosonic and/or fermionic particle species which are annihilated by the operators $\hat{\psi}_{\alpha}, \,\, \alpha=1,...,q$. These operators satisfy (anti)commutation relations 

\begin{eqnarray}
\hat{\psi}_{\alpha}(x) \hat{\psi}_{\beta}(x') - \eta_{\alpha \beta} \hat{\psi}_{\beta}(x')\hat{\psi}_{\alpha}(x) &=& 0 \\
\hat{\psi}_{\alpha}(x) \hat{\psi}^{\dagger}_{\beta}(x') - \eta_{\alpha \beta} \hat{\psi}_{\beta}^{\dagger}(x')\hat{\psi}_{\alpha}(x) &=& \delta_{\alpha \beta}\delta(x-x')
\end{eqnarray}
where $\eta_{\alpha \beta} = +1$ if at least one of the fields
$\alpha$ or $\beta$ is bosonic and $\eta_{\alpha \beta} = -1$ if both
fields are of fermionic nature. 

The $q$-species cMPS state is defined as \citep{Haegeman}
\begin{equation}\label{q-cMPS}
\vert \chi \rangle = {\rm tr_{aux}}\lbrace B \mathcal{P} {\rm e}^{
  \int_0^L {\rm d}z \, \tilde{Q}(z)\otimes \mathbb{1} + \sum_{\alpha =
    1}^q \tilde{R}_{\alpha}(z) \otimes
  \hat{\psi}_{\alpha}^{\dagger}(z)} \rbrace \vert \Omega \rangle \; , 
\end{equation}
here, matrices $\tilde{R}_{\alpha}$ have been introduced for each one
of the fields $\hat{\psi}_{\alpha}$ and a single Hamiltonian
$\tilde{K}$ for the auxiliary system. We will employ the tilde
notation for the variational parameters of the multi-species cMPS
state to differentiate them from their single field counterparts,
cf. Eq. \eqref{cMPS-def}. The matrix $\tilde{Q}$ is now defined as
\begin{equation}
\tilde{Q}(x)=-i\tilde{K}(x) - \frac{1}{2}\sum_{\alpha=1}^q \tilde{R}_{\alpha}^{\dagger}(x) \tilde{R}_{\alpha}(x)
\end{equation}

At difference with the single field case, a regularity condition must be imposed on the $\tilde{R}_{\alpha}$ matrices in order that the expectation value of the non-relativistic kinetic energy, as computed with (\ref{q-cMPS}), will not become divergent. This condition reads 
\begin{equation}\label{comm}
\tilde{R}_{\alpha}(x) \tilde{R}_{\beta}(x) - \eta_{\alpha \beta} \tilde{R}_{\beta}(x) \tilde{R}_{\alpha}(x) = 0
\end{equation}
In other words, the matrices $\tilde{R}_{\alpha}$ inherit the (anti)commutation relation of their corresponding fields. With these ideas in mind we can extend the operational rules for computing expectation values with cMPS. For example, 
\begin{equation}
\langle \chi \vert \chi \rangle = {\rm tr} \lbrace {\rm e}^{\tilde{T} L} \rbrace
\; ,
\end{equation}
where the transfer operator (\ref{transfer1}) has been generalized to 
\begin{equation}\label{newT}
\tilde{T} = \tilde{Q} \otimes \mathbb{1} + \mathbb{1} \otimes \tilde{Q}^* + \sum_{\alpha = 1}^q \tilde{R}_{\alpha} \otimes \tilde{R}_{\alpha}^*
\end{equation}
and translational invariance has been assumed for simplicity.

Special care must be taken into account for systems where two or more
fermionic species coexist.
Let us discuss correlators like $\langle
\hat{\psi}_{\alpha}^{\dagger}(x) \hat{\psi}_{\beta}(y)
\rangle$. 
Expanding the path-ordered exponential in (\ref{q-cMPS}), which acts on the vacuum
$\vert \Omega \rangle$ of the field theory, and taking the annihilation
operators to the right (normal ordering prescription) we obtain \cite{Haegeman}
\begin{align}
\langle \hat{\psi}_{\alpha}^{\dagger}(x) \hat{\psi}_{\beta}(y) \rangle
=   {\rm tr} \lbrace 
&
{\rm e}^{\tilde{T} y}
(\tilde{R}_{\beta} \otimes \mathbb{1}) {\rm e}^{\tilde{T}_{\alpha}
  (x-y)} 
\nonumber\\ 
& \times (\mathbb{1} \otimes \tilde{R}_{\alpha}^*) {\rm e}^{\tilde{T}(L-x)} \rbrace 
\end{align}
($x > y$) where the generalized transfer operator $\tilde{T}_{\alpha}$ deals with the exchange statistics
\begin{eqnarray}
\tilde{T}_{\alpha} &=& \tilde{Q} \otimes \mathbb{1} + \mathbb{1} \otimes \tilde{Q}^* + \sum_{\beta = 1}^q \eta_{\alpha \beta} \tilde{R}_{\beta} \otimes \tilde{R}_{\beta}^* 
\end{eqnarray}
For the case of bosonic systems, $\tilde{T}_{\alpha} = \tilde{T}$.
The transfer operator $\tilde{T}$ governs the evolution of states in the ancillary space. Similarly to the single field case, this evolution can be mapped to a dissipative dynamics corresponding to the following Linblad quantum master equation
\begin{equation}\label{Lind2}
\frac{d \tilde{\varrho} (z)} {d z}
= 
-i [\tilde{K}, \tilde{\varrho}(z)]
+
\sum_{\alpha = 1}^q \tilde{R}_{\alpha} \tilde{\varrho} \tilde{R}_{\alpha}^\dagger - \frac{1}{2} [ \tilde{R}_{\alpha}^\dagger \tilde{R}_{\alpha}, \tilde{\varrho}(z)]_+
\end{equation}
Thus, we have again the picture of the ancilla coupled to a bath (the fields)
by means of the operators $\tilde{R}_{\alpha}$. 

Consider now the case of two bosonic fields $\hat{\psi}_1$ and $\hat{\psi}_2$. We are interested in studying how the matrices ($\tilde{R}_{\alpha}$ and $\tilde{K}$), which define the cMPS state in this two-species system, can be constructed from the matrices which characterize a single field. The simplest scenario considers two uncoupled fields. We have seen how the problem of computing expectation values in the ground state can be reduced to a dissipative dynamics going on in the auxiliary space - where the state of the total auxiliary system is described in terms of the density matrix $\tilde{\rho}$. In the absence of a coupling between the fields, we should be able to recover our single field solutions. This is nothing but to demand the  density matrix to be separable, that is, $\tilde{\varrho} = \varrho_1 \otimes \varrho_2$. Both fields do not need to be identical, therefore, each of them will have associated a different set of matrices $R_{\alpha}$ and $K_{\alpha}$ which act on the corresponding auxiliary space $\mathcal{A}_{\alpha}$. For simplicity, we assume that both  $\mathcal{A}_1$ and $\mathcal{A}_2$  have the same bond dimension $D_1 = D_2= D$. The total auxiliary space for the two fields will be the tensor product of the individual spaces $\tilde{\mathcal{A}} = \mathcal{A}_1 \otimes \mathcal{A}_2$. Due to the tensor product structure, the bond dimension of the total auxiliary space is now $\tilde{D} = D^2$. The ancillas evolve independently according to the total Hamiltonian
\begin{equation}\label{K-coupled}
\tilde{K} = K_1 \otimes \mathbb{1} + \mathbb{1} \otimes K_2 
\end{equation}
Similarly, each auxiliary system will couple to its quantum field by means of the matrices $R_{\alpha}$. The extension of these to the product space is
\begin{eqnarray}\label{R1}
\tilde{R}_1 = R_1 \otimes \mathbb{1}
\end{eqnarray}
\begin{equation}\label{R2}
\tilde{R}_2 = \mathbb{1} \otimes R_2
\end{equation}
Notice that the matrices $\tilde{R}_{\alpha}$ satisfy the bosonic
commutation relation $[\tilde{R}_1,\tilde{R}_2] = 0$ as it is demanded
for a multi-species system (\ref{comm}). As desired, our construction
let us recover the results for single fields. For instance, $\langle
\hat{\psi}_1^{\dagger} \hat{\psi}_1 \rangle = {\rm tr} (\tilde{\rho}^*
\tilde{R}_1^{\dagger} \tilde{R}_1) = {\rm tr}(\rho_1^* R_1^{\dagger}
R_1) {\rm tr}\rho_2^* = {\rm tr}(\rho_1^* R_1^{\dagger} R_1)$ (as
${\rm tr}\rho_2^* = 1$ for a density matrix). 

How is this picture modified in the presence of a coupling between $\hat{\psi}_1$ and $\hat{\psi}_2$?
An arbitrary operator $\tilde{C}$, mapping $\tilde{\mathcal{A}}$ into
itself, can be represented as  $\tilde{C} = \sum_i c_i A_i \otimes
B_i$, where $A_i$ acts on $\mathcal{A}_1$ and $B_i$  acts on
$\mathcal{A}_2$. Therefore, this defines the most general structure
for the matrices $\tilde{R}_{\alpha}$ and $\tilde{K}$. 
Those general matrices must satisfy the regularity conditions
(\ref{comm}), which complicates their construction.

A possible solution is the following.  We use the intuitive
interpretation for the cMPS in terms of a system-bath, see Fig.
\ref{fig:sketch}(b) and the discussion on the holographic
interpretation below Eq. \eqref{Lind}.  Starting from the decoupled
solution (\ref{K-coupled}), (\ref{R1}) and (\ref{R2}), we switch on the
coupling adiabatically and expect that our solutions will start to
modify. This is depicted schematically in
Fig. \ref{fig:sketch}(b). Here, as one introduces the coupling between
the physical fields, the individual auxiliary spaces will also start
to  \emph {interact}. 
Inspired by this procedure, 
we propose the following construction in the presence of a coupling.
 First of all, the matrices $\tilde{R}_1$ and $\tilde{R}_2$ will
 continue to be described by (\ref{R1}) and (\ref{R2})
 respectively. In this way, we guarantee that they commute, satisfying
 \eqref{comm} trivially. In order to render the state non-separable,
 the matrix $\tilde{K}$ is written in a general way but  containing
 the uncoupled solution as a limit (\ref{K-coupled}). This is done as follows
\begin{equation}\label{Kfull}
\tilde{K} = K_1 \otimes \mathbb{1} + \mathbb{1} \otimes K_2 + \sum_{p=0}^{P} Z_1^{(p)} \otimes Z_2^{(p)} 
\end{equation}
where $Z_1^{(0)} =Z_2^{(0)} = 0$. In order to keep $\tilde{K}$ Hermitean, we will demand that the matrices $Z_i^{(p)}$ are Hermitean too. The number $P$ of pairs of $Z$ matrices is arbitrary. In principle, we will expect it to grow with the strength of the coupling.

We have seen for the single field case, that the cMPS {\it ansatz} is able to map the properties of a continuous one-dimensional field theory by means of $2 D^2$ variational parameters (with of course, a relatively small bond dimension $D$). Doubling the number of fields, as well as introducing $P$ pairs of the already defined $Z$ matrices, increases the total number of variational parameters to $(4+2P) D^2$.

%%%%%%%%%%%%%%%%%%%%%%%%%%%%%%%%%%%%%%%%
%%%%%%%%%%%%%%%% results for coupled  %%%%%%%%%%%%%%%%%
%%%%%%%%%%%%%%%%%%%%%%%%%%%%%%%%%%%%%%%%
\section{Two-species bosonic system}
\label{sec:coupledLL}

The system we have in mind to test the cMPS method for coupled fields
is a two-component bosonic system. Binary systems of this kind (as
well as bosonic + fermionic mixtures) are Luttinger liquids with a
rich phase diagram \cite{Cazalilla2, Mathey}. We will consider two
Lieb-Liniger gases with a density density coupling. This is described
by the following Hamiltonian:
\begin{align}
\label{H-coupled}
\hat{H} = 
\frac{1}{2 M} \sum_{\alpha =1}^2
&
 \int_0^L {\rm d} x \,
\partial_x \hat{\psi}_{\alpha}^{\dagger}(x) \partial_x
\hat{\psi}_{\alpha} (x) \nonumber\\
+ c \sum_{\alpha =1}^2
&
 \int_0^L {\rm d} x \, 
\big ( \hat{\psi}_{\alpha}^{\dagger}(x) \big )^2
\big ( \hat{\psi}_{\alpha}(x) \big )^2
%\hat{\psi}_{\alpha}^{\dagger}(x)
% \hat{\psi}_{\alpha}^{\dagger}(x) \hat{\psi}_{\alpha}(x)
% \hat{\psi}_{\alpha}(x)  
\nonumber\\
+  g & \int_0^L {\rm d}x \, \hat{\rho}_1(x) \hat{\rho}_2(x)
\end{align}

In order to obtain the low-energy behaviour of this model we will use
the bosonization technique introduced in Sect. \ref{sec:boso}. As
already explained, this consists in rewriting the bosonic fields in
terms of the collective fields $\hat{\theta}_{\alpha}$ and
$\hat{\phi}_{\alpha}$ which characterize the bosonic low-energy
excitations.  Hamiltonian (\ref{H-coupled}) conserves the individual particle densities, $[\hat H, \hat
\rho_\alpha]=[\hat H, \hat \rho]=0$ ($\alpha=1,2$).
 Therefore, we can fix these two densities as minimization constraints in the cMPS procedure.  In Eq. (\ref{density-expansion}), we have considered the lowest order term in a harmonic expansion of the density operator. A more careful treatment\cite{giamarchi} shows that,  the correct expansion for the density operator in terms of the field $\hat{\phi}_{\alpha}$ is of the form $\hat{\rho}_{\alpha}(x)=[\rho_{0 \alpha} - \partial_x \hat{\phi}_{\alpha}(x) /\pi]\sum_p{\rm e}^{i2p(\pi \rho_{0 \alpha} x - \hat{\phi}_{\alpha}(x))}$. Our former simplification is justified due to the fact that, at long distances (low-energies), the phase terms oscillate very fast and will average to zero upon integration. In performing the bosonization, we must retain the most dominant terms at low-energies. For the case of our inter-species coupling, this supposes to consider also the first harmonic $p=1$. This leads, at low temperatures, to a coupling contribution of the form: $1/2\pi \int {\rm d}x \,[ 2g_x \partial_x \hat{\phi}_1 \partial \hat{\phi}_2 + g_c \cos(2(\hat{\phi}_1 - \hat{\phi}_2) + \pi \delta x)]$ (with $\delta = \rho_{01} - \rho_{02}$). Of particular interest for us will be the case of equal filling $\rho_{01} = \rho_{02}$ ($\delta = 0$). Species $1$ and $2$ in the low-energy effective Hamiltonian can be decoupled by introducing the normal modes  $\hat{\phi}_+ = 1/\sqrt{2} (\hat{\phi}_1 + \hat{\phi}_2)$ and $\hat{\phi}_- = 1/\sqrt{2} (\hat{\phi}_1 - \hat{\phi}_2)$. In terms of these we have that the low-energy excitations of (\ref{H-coupled}) can be described by the effective Hamiltonian
\begin{align}
\nonumber
\hat{H}_{\rm eff} = \frac{1}{2\pi} \int {\rm d}x \, 
\Big [
& \sum_{\nu=\pm} v_\nu \left(
    (\partial_x \hat{\phi}_\nu)^2 + \frac{1}{K_\nu} (\partial_x
    \hat{\theta}_\nu)^2 \right) 
\\ 
& + g_c \cos(\sqrt{8} \hat{\phi}_-) \Big ]
\label{H-eff-coupled}
\end{align}
Similarly to the single field case, the Luttinger parameters $v_{\pm}$ and $K_{\pm}$ can be related to the ground state energy density $e_0(\rho_{+}, \rho_{-})$ (as a function of the normal densities) of Hamiltonian (\ref{H-coupled}).
\begin{align}
\label{vpm}
v_{\pm}^2 &= \frac{2 \rho_{0\pm}}{M} \left. \frac{\partial^2 e_0}{\partial \rho_{\pm}^2} \right\vert_{\rho_{\pm} = \rho_{0\pm}} 
\\
\label{Kpm}
K_{\pm}^2 &= \frac{\pi^2 \rho_{0\pm}}{2 M} \left( \left. \frac{\partial^2 e_0}{\partial \rho_{\pm}^2} \right\vert_{\rho_{\pm} = \rho_{0\pm}} \right)^{-1}
\end{align}

\begin{figure}[t]
\includegraphics[width=0.98\columnwidth]{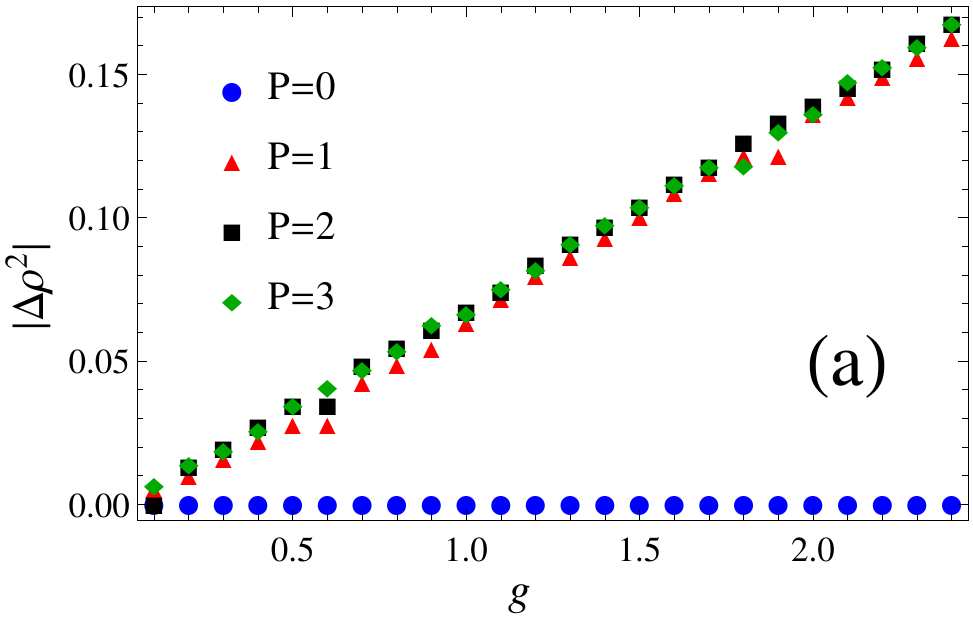}
\includegraphics[width=0.98\columnwidth]{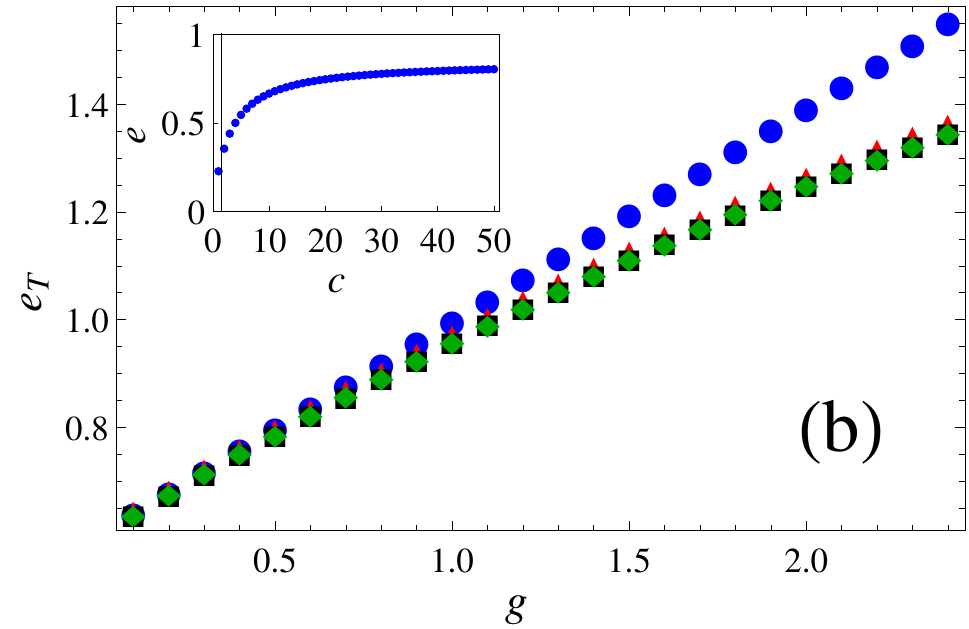}
 \caption{(color online) (a) Ground state density-density correlations and (b) total energy density for Hamiltonian (\ref{H-coupled}). Both as a function of the interspecies coupling strength $g$ for equal densities $\rho_{01} = \rho_{02} = 0.63$ and $c = 1.5$. The ancilla space for each field has bond dimension $D$ equal to $6$. We couple the ancillas by means of $P$ pairs of $Z$ matrices. Results are shown for $P=0$ (circles), $P=1$ (triangles), $P=2$ (squares) and $P=3$ (diamonds). Inset: ground state energy density for a single Lieb-Liniger chain as a function of $c$ ($D=6$). The vertical line denotes the value of $c$ at which we are coupling two fields.}
 \label{fig:corr-energy-coupled}
 \end{figure}

 Coupled species have been thoroughly studied \cite{Cazalilla2,
   Mathey}.
In this work we study coupled bosonic species described by \eqref{H-coupled}.
The range of parameters considered coincides with the one in Refs.
\onlinecite{Kleine, Kleine-thesis} where a DMRG study is reported, hence
a direct comparison is possible. 

In Fig. \ref{fig:corr-energy-coupled} (a) we plot the ground state
density-density correlations
 $
| \Delta \rho^2 | =\vert \langle \hat{\rho}_1(x) \hat{\rho}_2(x)
\rangle - \langle \hat{\rho}_1(x) \rangle \langle \hat{\rho}_2(x)
\rangle \vert$ as a function of the interspecies coupling $g$ for
different values of $P$. Both, the repulsion strength $c$ and the bond
dimension $D$ are kept fixed ($c=1.5$ and $D=6$). As expected, $P=0$
renders the state separable and no correlations are observed.
Making $P \neq 0$ the correlations between the two fields build up.
They grow with the coupling strength.  In this range of
parameters, $P=2$ seems to be sufficient for account with the
physics. 

The ground state energy density as a function of $g$ is shown in
Fig. \ref{fig:corr-energy-coupled} (b). Only the last term of
(\ref{H-coupled}) depends on $g$, therefore, (for a fixed value of
$c$) the first two terms yield a constant contribution. 
The case $P=0$ yields a mean field treatment where the interaction is
replaced by $g 
 \langle \hat{\rho1}  \rangle \langle \hat{\rho_2}  \rangle$.
It was already mentioned that $[\hat H, \hat \rho_1 ] = [\hat H, \hat
\rho_2 ]=0$.  Thus, with $P=0$, the energy as a function of $g$ has a
linear dependence with slope $\rho_{01} \rho_{02}  $.
Including quantum correlations ($P \neq0$) the energy is no longer a
linear function of the coupling,  as seen in
\ref{fig:corr-energy-coupled} (b).

\begin{figure}[t]
\includegraphics[width=0.98\columnwidth]{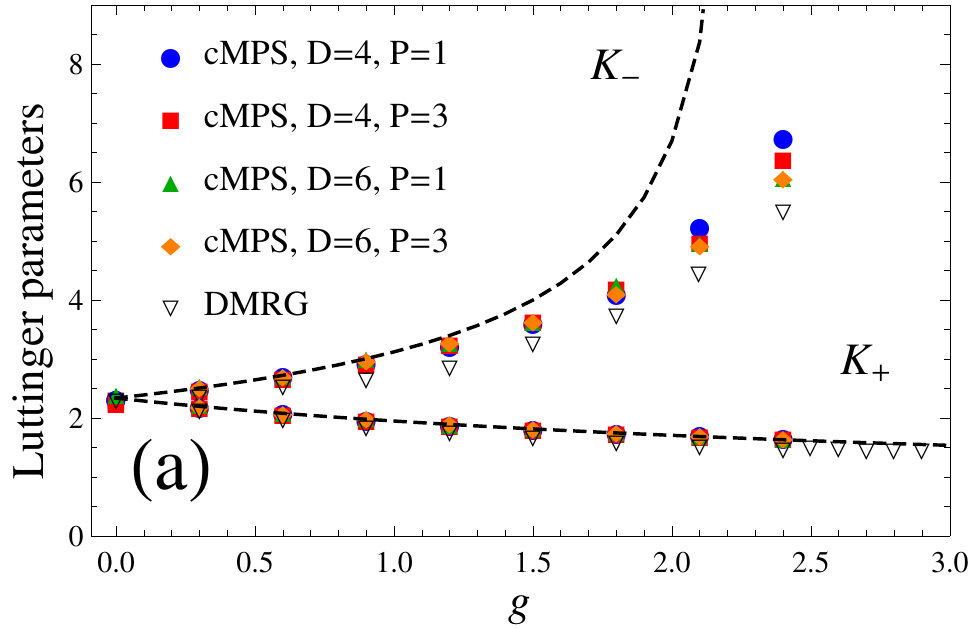}
\includegraphics[width=0.98\columnwidth]{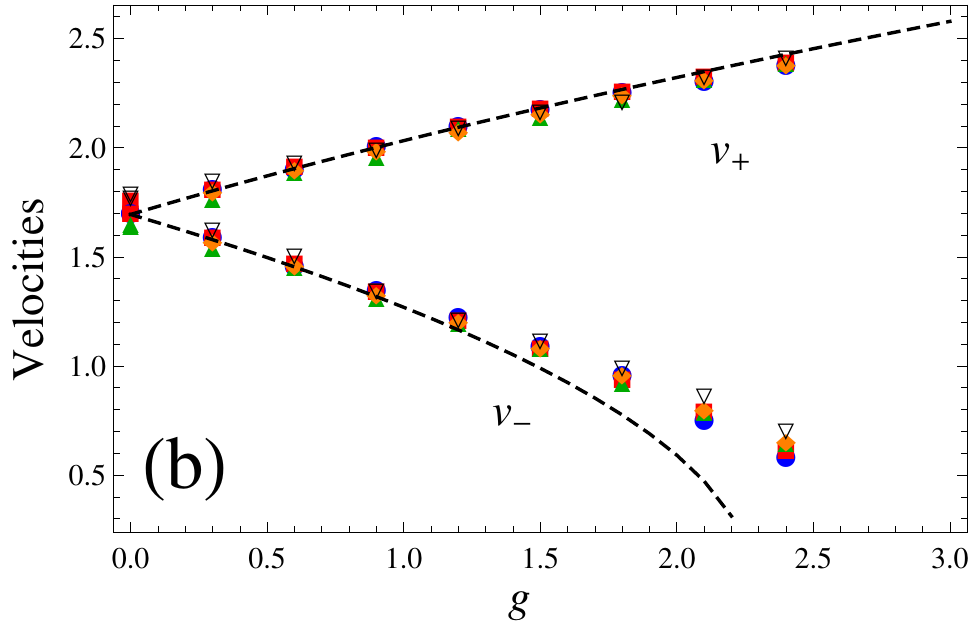}
 \caption{(color online) Charge(+) and spin(-) Luttinger parameters  as a function of the coupling $g$ between fields. Here we show a comparison between a weak-coupling approximation (dashed line), DMRG \cite{Kleine} and our results implementing cMPS for coupled fields with different bond dimensions and values of the parameter $P$ for fixed $c=1.5$. (a) $K_{\pm}$ (b) velocities $v_{\pm}$.}
 \label{fig:Luttinger-coupled}
 \end{figure}

As it was already discussed, the low-energy description of our model is 
characterized by the Luttinger parameters \eqref{vpm} and \eqref{Kpm}.
In particular, the difference in {\it normal velocities} $v_\pm$ yields
the charge-spin separation - a typical experimental characteristic in
mixtures.
In Fig. \ref{fig:Luttinger-coupled} we compare for both the $v_\pm$
and $K_\pm$ our cMPS results with the DMRG values extracted from
Refs. \onlinecite{Kleine, Kleine-thesis}. Let us remark the excellent
agreement in $K_+$ and $v_+$ and the minor discrepancy in $K_-$ or
$v_-$. The very small differences could be attributed to a
number of issues: small bond dimension in the cMPS ($D=6$) to be compared
with the DMRG (several hundreds), or the fact that the DMRG theory is
discretized an the cMPS is fully continuous. 

%%%%%%%%%%%%%%%%%%%%%%%%%%%%%%%%%%%%%%%%
%%%%%%%%%%%%%%%% discussion  %%%%%%%%%%%%%%%%%
%%%%%%%%%%%%%%%%%%%%%%%%%%%%%%%%%%%%%%%%
\section{Quantum simulation of coupled cMPS}
\label{sec:disc}

\begin{figure}[t]
\includegraphics[width=0.98\columnwidth]{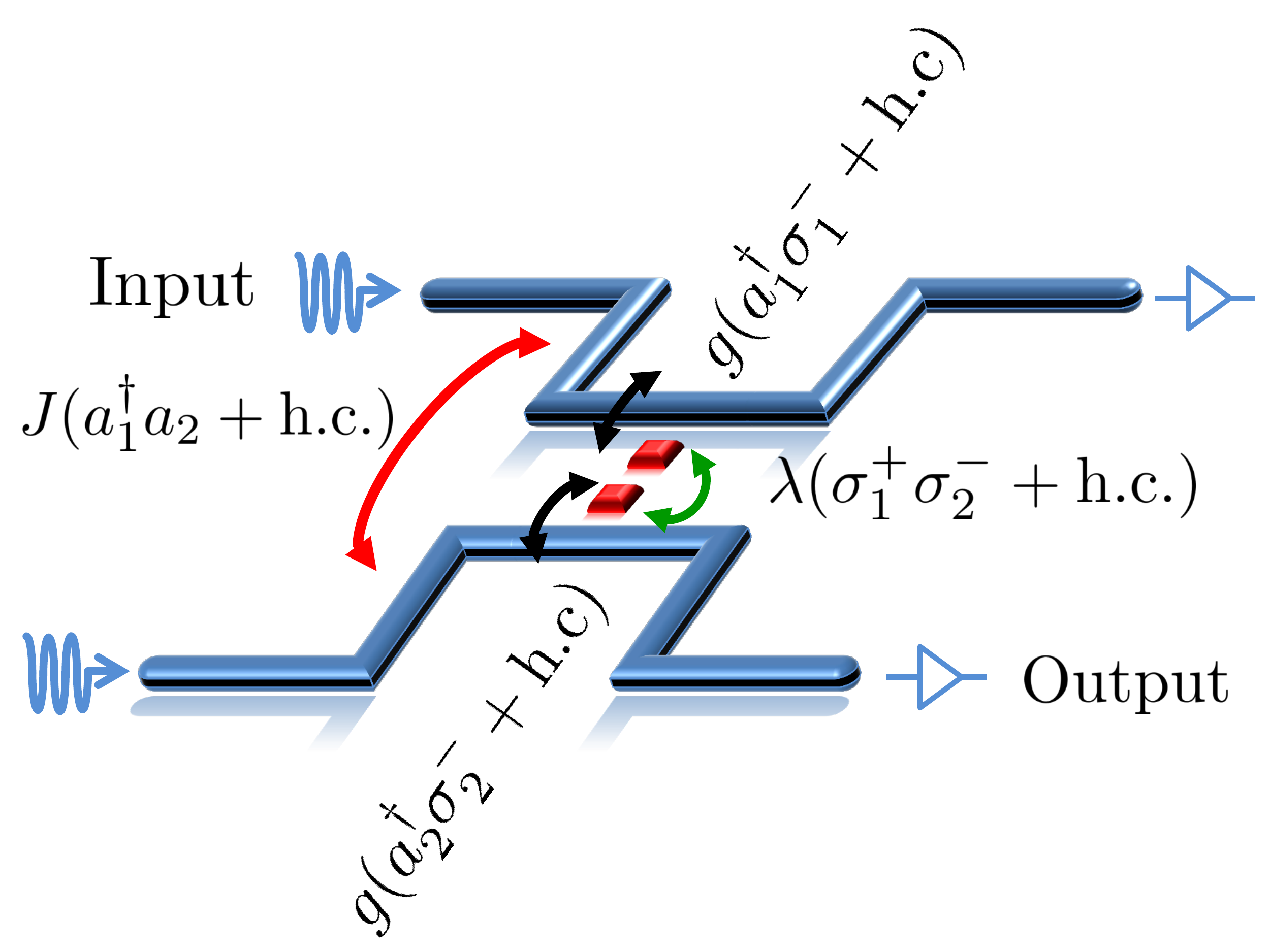}
 \caption{(color online) Possible circuit QED implementation for the quantum simulation of a cMPS for two coupled fields. The ancillas consist on two cavity-qubit setups and the fields are the input and output of the EM field. The Hamiltonian of the cavity-qubit setups simulates $\tilde{K}$ and the cavity operators couple to the external field, that is, they correspond to the matrices $\tilde{R}_{\alpha}$.}
 \label{fig:q-sim}
 \end{figure}

There exist two approaches towards the quantum simulation of
continuous or discrete field theories \cite{Nori}. The conventional one consists
on taking a flexible quantum system, such as a Bose-Einstein
condensate, ultracold atoms in an optical lattice or a superconductor,
and working with it to implement the full field theory, or an
approximate version of it, in the experiment. This ``analogue''
quantum simulator therefore evolves and equilibrates as the original
model dictates and all observables may be directly studied on the
experiment itself.

A second possibility for quantum simulation arises from the physical
interpretation of cMPS. The idea is that there exists a mapping
between a continuous Matrix Product state and a physical process
operating on a small quantum mechanical object. This mapping between
states and channels was already evidenced for discrete
MPS\cite{Schon2005,Schon2007} and has been recently generalized for
cMPS\cite{Barrett2013}, by means of their physical interpretation in
terms of a system (the ancilla) coupled to a bath (the field). The
beauty of this mapping is that it is quite general and applies to a
variety of quantum optical systems. The prototypical system is an
atom-cavity setup (the system or ancilla in the language of this
paper) that interacts with external \emph {input} and \emph { output}
fields through the bath (the field in cMPS) in this case
the electromagnetic field.  However, any other quantum discrete system
coupled to an outer field, where different order correlations of the
latter can be measured, such as circuit QED \cite{Wallraff2004,You2011} would do
the job.

Let us now summarize the proposal in Ref. \onlinecite{Barrett2013}. 
The atom-cavity system is described through the well known
Jaynes-Cummings (JC) model, 
\begin{equation}
\hat{H}_{\rm ancilla} = \hat{H}_{\rm JC}= \Omega \hat{a}^\dagger \hat{a} + \epsilon \hat{\sigma}^+
\hat{\sigma}^-+ g (\hat{a}^\dagger \hat{\sigma}^- + {\rm h.c} )
\end{equation}
here $\hat{a}$ ($\hat{a}^\dagger$) are bosonic annihilation (creation) operators
describing the main stationary mode of a cavity. The atom (with two relevant states splitted by
$\epsilon$) is coupled to this fundamental mode of $\Omega$-frequency
with a strength $g$. The $\hat{\sigma}^-$ ($\hat{\sigma}^+$) are lowering
(raising) operators for the two-level system.  
The atom-cavity is coupled to an EM-environment, that in second
quantization is given by the free Hamiltonian, $\hat{H}_{\rm EM} =
\int {\rm d}\omega \, \omega \, \hat{b}^{\dagger}(\omega) \hat{b}(\omega)
$.
Taking an interaction picture with respect to the EM field, the system-bath (ancilla-field) coupling can be written as
\begin{equation}
\hat{H}_{\rm coupling}(t) = \sqrt{\frac{\kappa}{2\pi}} \int {\rm d} \omega \, \hat{a}^\dagger \hat{b}(\omega) {\rm e}^{-i\omega t} +
{\rm h.c.}
\end{equation}
Limiting the integration region to frequencies $\omega$ near $\Omega$ we can safely assume the RWA. Also, assuming a point-like interaction in space, the coupling function is flat in momentum.
It is customary to introduce the time dependent operators $\hat{E}^+(t) = i/\sqrt{2\pi} \int {\rm d}\omega \, {\rm e}^{-i \omega t} \hat{b}(\omega)$ and Hermitian conjugate. They correspond to the electric field components of the EM field. In this way, we can finally write the total Hamiltonian as
\begin{equation}\label{HJC-EM}
\hat{H}(t)= \hat{H}_{\rm ancilla} + i\sqrt{\kappa} ( \hat{a} \otimes \hat{E}^-(t) - {\rm h.c.})
\end{equation}
The electric field operators can, in turn, be decomposed into {\it in}-{\it out} components\cite{Gardiner1985}. The {\it in} component corresponds to the field that impinges on the  system while the {\it out} component consists of a reflected part plus a radiated one due to the interaction of the EM field with the system.  
If we take the {\it in} state of the EM field to be the vacuum, it can be shown\cite{Gardiner-Zoller} that the evolution governed by (\ref{HJC-EM}) can be reduced to that of the non-Hermitian Hamiltonian
\begin{equation}
-i\hat{H}_{\rm eff}(t) = -i \hat{H}_{\rm ancilla} - \frac{1}{2}\kappa \hat{a}^{\dagger} \hat{a} + \sqrt{\kappa} \hat{a} \otimes \hat{E}^{-}(t)
\end{equation}
This is the same kind of evolution which generates the cMPS {\it ansatz} (see Eqs. (\ref{cMPS-def}) and (\ref{Q-single})) once we trace over the degrees of freedom of the ancilla. We thus make the following identification: 
\begin{equation}
R = \sqrt{\kappa} \hat{a} 
\qquad
K = \hat{H}_{\rm ancilla}(\Omega,\epsilon,g)
\end{equation}
While we do not have control over $R$, we can modify the variational parameter
$K$ by properly
tuning the couplings ($\Omega$,$\epsilon$,$g$) of the cavity-atom
system. The continuous field $\hat{\psi}(x)$ will map into the output
field operators of the electromagnetic field: $\hat{\psi}(x) =
\hat{E}^{+}(t)$. Being the EM field in a cMPS state,
computing expectation values of operators will translate into
measuring correlations of the EM field itself, {\it i.e.}, measuring
the normalized correlation functions $g^{(1)}(t,t')$, $g^{(2)}(t,t')$
\begin{eqnarray}
g^{(1)}(t,t') &=& \frac{\langle \hat{E}^-(t) \hat{E}^+(t')  \rangle}{\sqrt{\langle \hat{E}^-(t) \hat{E}^+(t) \rangle  \langle \hat{E}^-(t') \hat{E}^+(t')  \rangle}}  \\
g^{(2)}(t,t') &=& \frac{\langle \hat{E}^-(t) \hat{E}^-(t') \hat{E}^+(t) \hat{E}^+(t')  \rangle}{\langle \hat{E}^-(t) \hat{E}^+(t) \rangle  \langle \hat{E}^-(t') \hat{E}^+(t')  \rangle}
\end{eqnarray}
and higher orders depending on the model we wish to simulate. Following our previous identification, the correlators $\langle \hat{E}^-(t) \hat{E}^+(t') \rangle$ and $\langle \hat{E}^-(t) \hat{E}^-(t') \hat{E}^+(t) \hat{E}^+(t')  \rangle$ map to $\langle \hat{\psi}^{\dagger}(x) \hat{\psi}(x') \rangle$ and $\langle \hat{\psi}^{\dagger}(x) \hat{\psi}^{\dagger}(x') \hat{\psi}(x) \hat{\psi}(x')  \rangle$ respectively.
It was shown numerically that the atom-cavity setup could simulate the
Lieb-Liniger model giving correlations  acceptably well \cite{Barrett2013}.

With this work at hand, our proposal has also a natural realization.
In our case, we envision two superconducting cavities interacting
each one with one\cite{Sun2014, Mariantoni2008, Reuther2010} or several superconducting qubits (Fig. \ref{fig:q-sim}):
\begin{equation}
\hat{H}_{\rm sys} = \sum_{\alpha}   g \hat{a}_{\alpha}^{\dagger}
\hat{\sigma}_{\alpha}^- + J \hat{a}_{\alpha}^{\dagger}
\hat{a}_{\alpha +1 } + \lambda \hat{\sigma}_{\alpha}^+
\hat{\sigma}^-_{\alpha +1 } + {\rm h.c.} 
\end{equation}
That are coupled to different baths (different fields) through
\begin{equation}
\hat{H}_{\rm coupling}(t) = \sum_\alpha \sqrt{\kappa_\alpha} \int {\rm d} \omega \, \hat{a}_\alpha^\dagger \hat{b}_\alpha(\omega) {\rm e}^{-i \omega t} +
{\rm h.c.}
\end{equation}
Therefore in our case ($\alpha = 1,2$), the identifications are the following,
\begin{equation}\label{R-QS}
\tilde R_\alpha = \sqrt{\kappa_\alpha} \hat{a}_\alpha
\end{equation}
and 
\begin{equation}\label{Kcoupled-QS}
\widetilde K = \hat{H}_{\rm sys}
\end{equation}
with
\begin{equation}\label{Ksingle-QS}
K_\alpha = \hat{H}_{\rm JC, \alpha} = g \hat{a}_{\alpha}^{\dagger} \hat{\sigma}_{\alpha}^- + {\rm h.c.}
\end{equation}
and
\begin{equation}\label{Z1-QS}
Z_1^{(1)} \otimes Z_2^{(1)} + Z_1^{(2)} \otimes Z_2^{(2)} = J \hat{a}_1 \otimes \hat{a}^\dagger_2 + {\rm h.c.}
\end{equation}
and
\begin{equation}\label{Z2-QS}
Z_1^{(3)} \otimes Z_2^{(3)} + Z_1^{(4)} \otimes Z_2^{(4)} = \lambda \hat{\sigma}_1^+ \otimes \hat{\sigma}_2^- + {\rm h.c.}
\end{equation}

Note that in Sect.\ref{sec:extension} we demanded that the matrices $Z$ should be Hermitian. Eqs. (\ref{Z1-QS}) and (\ref{Z2-QS}) can always be brought into a sum of tensor products of Hermitian operators. The former equations are of the form $C = A \otimes B^{\dagger} + A^{\dagger} \otimes B$. We can split any operator in terms of its Hermitian components. In the case of $A$, the decomposition reads: $A = A_r + i A_i$ (and similarly for $B$). Here, $A_r = 1/2 (A^{\dagger} + A)$ and $A_i = i/2 (A^{\dagger} - A)$. It is straightforward to show that $C$ can be rewritten as: $C = 2A_r \otimes B_r + 2A_i \otimes B_i $.  

Finally, as for the single field, EM field correlations need to be computed. In addition, cross-correlations, for instance, $\langle \hat{E}_i^-(t) \hat{E}_j^+(t') \rangle$ will be necessary. In
circuit QED this is possible as reported in the literature
\cite{Eichler2011, DiCandia2014, Menzel2010, Bozyigit2010, Bozyigit2011}.  

%%%%
\section{Summary and conclusions}

In this work we have proposed an extension of continuous Matrix
Product States (cMPS) to study the ground state properties of 1D
coupled fields.  Our treatment has been confronted to previous DMRG
numerical results, showing good convergence properties even for
moderately large coupling strengths.  Finally, we have discussed how
it could be possible to realize computations for coupled fields using
a quantum simulator to implement the cMPS ansatz and optimizing over
the \emph {ansatz} parameters \cite{Barrett2013}. We believe that
extensions of this ansatz, together with new ideas on time evolution
and the study of quasiparticle excitations\ \cite{Haegeman2013,Haegeman2013b} can
provide a valuable insight on existing experiments with 1D atomic
Bose-Einstein condensates\ \cite{Langen2013,Steffens2014b}.
\\

\section*{Acknowledgements}

We acknowledge support from the Spanish DGICYT under
Projects No. FIS2011-25167 and FIS2012-33022, by the
Aragon (Grupo FENOL) and the EU Project PROMISCE. The authors would also like
to  acknowledge the Centro de Ciencias de Benasque Pedro 
Pascual for its hospitality.
\\\\

\bibliographystyle{apsrev4-1}
\bibliography{cMPS-method}

\end{document}